\documentclass[%
 reprint,
 showpacs,preprintnumbers,
 amsmath,amssymb,
 aps,
 prb,
]{revtex4-1}
\usepackage{dcolumn}
\usepackage{subfigure}
\usepackage{graphicx,color}
\usepackage{mathrsfs}

\makeatletter
\def\btt#1{\texttt{\@backslashchar#1}}
\DeclareRobustCommand\bblash{\btt{\@backslashchar}} \makeatother

\begin{document}

\title{Optical conductivity of the threefold Hopf semimetal}

\author{Tetsuro Habe}
\affiliation{Nagamori Institute of Actuators, Kyoto University of Advanced Science, Kyoto 615-8577, Japan}

\date{\today}

\begin{abstract}
A multifold Hopf semimetal is a topological point node semimetal possessing an anisotropy in the internal electronic structure, e.g., the dipole structure of the Berry curvature.
In this paper, the unique features of threefold Hopf semimetals in terms of the optical conductivity are theoretically investigated with a minimal theoretical model by using linear response theory for a linearly polarized photon.
The frequency spectrum of the optical conductivity shows an anisotropic dependence on the polarization angle of the incident photon even if the electronic band structure is completely isotropic.
The longitudinal optical conductivity linearly depends on the photon frequency and possesses step-like changes in the frequency spectrum.
The anisotropic electronic structure has a varying the number of steps with the orientation of the photon polarization axis.
We reveal that the anisotropy is attributed to symmetries preserving the point node in threefold Hopf semimetals.
The linearly polarized photon also induces a Hall current but it vanishes with the photon polarization axis parallel to the Berry dipole axis.
The numerical calculations show that these characteristic features can be observed even with a non-zero temperature and disorder.
\end{abstract}

\maketitle
\section{Introduction}
Topological nodal semimetals are characterized by electronic bands with point nodes or line nodes, on which two or more bands are touching.\cite{Murakami2007,Burkov2011,Young2012,Bradlyn2016}
Around the nodes, the electronic states present an unusual structure of the Berry curvature in the wave number space.
For instance, the vector field of the Berry curvature is the same as that produced by a monopole in Weyl semimetals.\cite{Murakami2007}
Recently, a novel type of topological nodal semimetal, the so-called Hopf semimetal, has been theoretically proposed and predicted to possess multifold point nodes at which three or more bands are attached.\cite{Graf2022}
Around the nodes, the dispersion of energy bands is flat or linear, and it is qualitatively the same as that of multifold point node semimetals\cite{Wieder2016,Bradlyn2016}.
However, the Berry curvature possesses a dipole structure different from the isotropic monopole in these multifold point-node semimetals\cite{Chang2018,Flicker2018}.
Since the anisotropic Berry curvature remains even with an isotropic band structure, the multifold Hopf semimetals possess internal anisotropy in electronic states.

The electronic properties of topological nodal semimetals have been investigated in terms of several measurements.
The optical conductivity represents the electronic transport properties associated with excited electrons by photon irradiation and has been investigated theoretically and experimentally for topological nodal semimetals.\cite{Xiao2016,Neubauer2016,Tabert2016,Carbotte2016,Mukherjee2017,Ahn2017,Barati2017,Schilling2017,Habe2018,Sanchez2019,Habe2019,Xu2020}
The frequency dependence of the optical conductivity describes the characteristic energy dispersion in topological nodal semimetals.
For instance, point node semimetals show a linear dependence on the frequency and line-node semimetals present a frequency independent flat spectrum.\cite{Xiao2016,Neubauer2016}
Moreover, recent work suggests that the internal structure of the wave function also affects the optical conductivity and can be observed as the variation of the spectrum.\cite{Habe2019}
In topological nodal semimetals, the low-energy excitation is associated only with the band structure around the nodes even if there are other Fermi pockets.
The optical conductivity is a useful observable for investigating the electronic property attributed to the electronic structure around the node in topological nodal semimetals.

In this paper, the optical conductivity of threefold Hopf semimetals is theoretically investigated, and the characteristic spectrum attributed to the unusual electronic structure is presented.
In Sec.\ref{sec_model}, a minimal model and the electronic structure are reviewed.
In Sec.\ref{sec_optical_conductivity}, the optical conductivity is presented analytically in the clean limit at zero temperature and numerically in the disordered case at non-zero temperature.
Moreover, the anisotropic behaviors in the spectrum are discussed in terms of symmetries of multifold Hopf semimetals.
The conclusion is given in Sec.\ref{sec_conclusion}.

\begin{figure}[htbp]
\begin{center}
 \includegraphics[width=80mm]{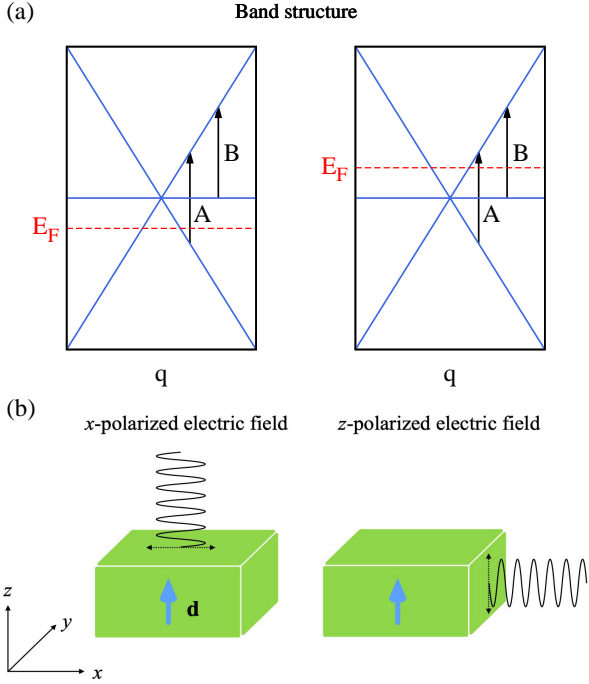}
\caption{Schematic band structure of a threefold Hopf semimetal in (a). The arrows indicate the optical excitation processes with a frequency of the photon.  The left and right panels show the cases with the Fermi energy below and above the point node, respectively.  In (b), schematics of the incident electric field with respect to the direction of dipole $\boldsymbol{d}$.
 }\label{fig_band}
\end{center}
\end{figure}
\section{Theoretical model}\label{sec_model}
In this section, the electronic structure of a threefold Hopf semimetal is briefly reviewed in terms of the electronic states and Berry curvature around the point node.
The theoretical analysis is performed by using a minimal model proposed in Ref.\ \onlinecite{Graf2022}.
Around the point node, the electronic states are described by
\begin{align}
H^\xi(\boldsymbol{q})=\hbar{v}\begin{pmatrix}
0&q^\xi_-&-iq_z\\
q^\xi_+&0&0\\
iq_z&0&0
\end{pmatrix},
\end{align}
where $\boldsymbol{q}$ is a wave vector with respect to the nodal point with $q^\xi_\pm=\xi q_x\pm iq_y$, and the isotropic speed $v$ is assumed.
Here, $\xi=\pm1$ indicates the direction of the dipole structure,  $\boldsymbol{d}=(0,0,\xi)$.
In previous work\cite{Graf2022}, two symmetries were proposed for the protection of the point node: axial rotation symmetry and chiral symmetry.
The axial rotation operator $L^d+\Sigma^d$ is defined with the angular momentum operator, $L^d=\boldsymbol{d}\cdot(i\nabla_{\boldsymbol{q}}\times\boldsymbol{q})$, projected on the $\boldsymbol{d}$-vector and the effective spin projection operator $\Sigma^d=(1/3)\mathrm{diag}[1,-2,1]$.
The symmetry for the rotation is given by
\begin{align}
[L^d+\Sigma^d,H^\xi(\boldsymbol{q})]=0.
\end{align}
The chiral operator for the minimal model is represented by $S=\mathrm{diag}[1,-1,-1]$ where chiral symmetry is given by $S^{-1}H^\xi(\boldsymbol{q})S=-H^\xi$.
In addition to the two symmetries, the mirror symmetry along the $\boldsymbol{d}$-vector,  the $z$-direction in the present case, is necessary for the point node to be protected.
For the minimal model, the mirror symmetry is represented by $M_z^\dagger H^\xi(q_x,q_y,-q_z)M_z=H^\xi(\boldsymbol{q})$ with $M_z=\mathrm{diag}[1,1,-1]$.
In practice, a matrix with a constant scalar $c$,
\begin{align}
H'=\begin{pmatrix}
0&0&c\\
0&0&0\\
c&0&0
\end{pmatrix},
\end{align}
removes the node without breaking the axial rotation symmetry and chiral symmetry.
In the other axes, there is no mirror symmetry because of the presence of axial mirror rotation symmetry.

The minimal model provides a characteristic band structure of threefold point node semimetals, i.e., one flat band ${\varepsilon}_{0,\boldsymbol{q}}=0$ and two cone-shaped bands,
\begin{align}
{\varepsilon}_{\pm,\boldsymbol{q}}=\pm \hbar v q,\label{eq_energy}
\end{align}
with $q=|\boldsymbol{q}|$, where these three bands are attached at $\boldsymbol{q}=0$ as shown in Fig.\ \ref{fig_band}. 
The electronic states are represented by scale-free wave functions,
\begin{align}
|\psi^\xi_{\pm,\boldsymbol{q}}\rangle=\frac{1}{\sqrt{2}q}\begin{pmatrix}
q\\
\pm q^\xi_+\\
\pm iq_z
\end{pmatrix},
\end{align}
and
\begin{align}
|\psi^\xi_{0,\boldsymbol{q}}\rangle=\frac{1}{q}\begin{pmatrix}
0\\
iq_z\\
q^\xi_-
\end{pmatrix},
\end{align}
where the subscript $\alpha$ of $\psi_{\alpha,\boldsymbol{q}}^\xi$ corresponds to the band index.
The simple representation enables us to analytically describe the fascinating electronic properties of threefold Hopf semimetals.

The Berry curvature is theoretically given by a Kubo-formula\cite{Yao2004},
\begin{align}
\Omega_{\alpha,\rho}(\boldsymbol{q})=-\epsilon_{\mu\nu\rho}\sum_{\alpha'\neq\alpha}\frac{2\mathrm{Im}[\langle\psi_{\alpha,\boldsymbol{q}}^\xi|v_\mu|\psi_{\alpha',\boldsymbol{q}}^\xi\rangle\langle\psi_{\alpha',\boldsymbol{q}}^\xi|v_\nu|\psi_{\alpha,\boldsymbol{q}}^\xi\rangle]}{|{\varepsilon}_{\alpha,\boldsymbol{q}}-{\varepsilon}_{\alpha',\boldsymbol{q}}|^2},
\end{align}
with Levi-Civita symbol $\epsilon_{\mu\nu\rho}$ in the orthogonal coordinate, where the velocity operator is defined by  $v_\mu=(1/\hbar)\partial H^\xi/\partial q_\mu$.
In the case of the threefold Hopf semimetal, the inter-band components of the velocity operator are scale-free and highly anisotropic,
\begin{align}
\begin{aligned}
\langle\psi_{0,\boldsymbol{q}}^\xi|\boldsymbol{v}|\psi_{\pm,\boldsymbol{q}}^\xi\rangle&=\frac{v}{\sqrt{2}q}\left(-i\xi{q_z},{q_z},i{q_\pm^\xi}\right)\\
\langle\psi_{+,\boldsymbol{q}}^\xi|\boldsymbol{v}|\psi_{-,\boldsymbol{q}}^\xi\rangle&=\frac{v}{q}\left(-i\xi q_y,iq_x,0\right).
\end{aligned}\label{eq_velocity}
\end{align}
The anisotropy leads to the dipole structure of the Berry curvature,
\begin{align}
\begin{aligned}
\boldsymbol{\Omega}_0(\boldsymbol{q})&=2\frac{\xi}{\hbar^2}(\boldsymbol{d}\cdot\boldsymbol{q})\frac{\boldsymbol{q}}{q^4}\\
\boldsymbol{\Omega}_\pm(\boldsymbol{q})&=-\frac{\xi}{\hbar^2}(\boldsymbol{d}\cdot\boldsymbol{q})\frac{\boldsymbol{q}}{q^4},
\end{aligned}
\end{align}
i.e., the Berry curvature vanishes in a plane of $q_z=0$.
The anisotropic Berry curvature shows that the internal anisotropy can be found in electronic properties even if the energy dispersion is completely isotropic.
In previous work\cite{Graf2022}, it was proposed that the anisotropy of the internal electronic structure can be confirmed by the Hall conductivity.
However, the band structure of a multifold nodal semimetal cannot be confirmed in the measurement simultaneously.
Moreover, the Hall conductivity is also affected by the presence of other nodes with the opposite $\boldsymbol{d}$-vector and/or conventional Fermi pockets.
On the other hand, the optical conductivity can represent both characteristics of internal anisotropy and a multifold nodal semimetal, and it does not vary even in the presence of other Fermi pockets.
The frequency dependence is related to the energy dispersion of the multifold nodal semimetal, and the different number of step-like structures can be found in the spectrum by changing the polarization axis of the photon.

\section{Optical conductivity}\label{sec_optical_conductivity}
The optical conductivity is a response function representing the induction of electric current by an oscillating electric field with a unique frequency, e.g. , the photo-induced electric current.
In this paper, the electric field is considered to be polarized along a unique direction for simulating a linearly polarized photon. 
The external field is represented by $E_\mu(t)=E^0_{\mu}\cos(\omega t)$ with the polarization axis in the $\mu$ direction and the frequency $\omega$.
In the presence of the oscillating field, electrons are excited by absorbing the photon energy corresponding to the frequency and the negligibly small change in wave number.
Since the excited electrons mainly carry the electric current, the frequency dependence of the optical conductivity characterizes the distribution of the excitation energy for electronic states in the wave number space.

In linear response theory, the induced electric current in the $\nu$ direction is represented by the superposition of those attributed to two complex fields, $E_{\pm\omega,\mu}(t)=E^0_{\mu}\exp[\mp i\omega t]$,
\begin{align}
J_\nu(t)=\frac{1}{2}\left(J_{\omega,\nu}(t)+J_{-\omega,\nu}(t)\right),
\end{align}
with 
\begin{align}
J_{\omega,\nu}(t)=-\int_{-\infty}^t{dt'}\ \mathrm{Tr}\left\{\rho_0[\hat{j}_\nu(t),\hat{j}_\mu(t')]\right\}\frac{E_{\omega,\mu}(t')}{\omega}.
\end{align}
By using the eigenstate $|m\rangle$, the optical conductivity $\sigma_{\nu\mu}(\omega)$ is given by
\begin{align}
\sigma_{\nu\mu}(\omega)=&J_{\omega,\nu}(t)/E_{\omega,\mu}(t)\nonumber\\
=&\sum_{{\varepsilon}_m\leq {\varepsilon}_n}\frac{ie^2}{\omega}\left(n_F({\varepsilon}_m)-n_F({\varepsilon}_n)\right)\nonumber\\
&\times\left\{\frac{v_\nu^{mn}v_\mu^{nm}}{\hbar\omega+{\varepsilon}_m-{\varepsilon}_n+i\delta}-\frac{(v_\nu^{mn}v_{\mu}^{nm})^\ast}{\hbar\omega-{\varepsilon}_m+{\varepsilon}_n+i\delta}\right\},\label{eq_response_function}
\end{align}
with a small imaginary factor $i\delta$, where $n_F(\varepsilon)$ and $v_\mu^{mn}=\langle m|v_\mu|n\rangle$ are Fermi distribution function and the matrix component of the velocity operator, respectively.
Here, the summation over the electronic states is performed under the condition of ${\varepsilon}_m\leq {\varepsilon}_n$.
The response function satisfies $\sigma_{\nu\mu}(-\omega)=\sigma_{\nu\mu}(\omega)^\ast$. 
Thus, the electric current can be represented by a real function with two oscillating functions,
\begin{align}
J_{\nu}(t)=\mathrm{Re}[\sigma_{\nu\mu}(\omega)]E_\mu^0\cos\omega t+\mathrm{Im}[\sigma_{\nu\mu}(\omega)]E_\mu^0\sin\omega t.
\end{align}
Since the external electric field oscillates with $\cos\omega t$, the first term represents the dissipative electric current, i.e., the observable.
Thus, the coefficient $\sigma^{(1)}_{\nu\mu}(\omega)=\mathrm{Re}[\sigma_{\nu\mu}(\omega)]$ gives the optical conductivity to the linearly polarized electromagnetic wave with a frequency $\omega$.

\subsection{Clean limit}
In the clean limit, the optical conductivity can be obtained by introducing an infinitesimal into $\delta$ in Eq. (\ref{eq_response_function}).
In this limit, the analytic representation is given by
\begin{align}
\sigma^{(1)}_{\nu\mu}(\omega)=&\frac{e^2}{\hbar}\sum_{\varepsilon_m\leq\varepsilon_n}\frac{\Delta n^{mn}_F}{\omega}\nonumber\\
&\times\left\{\mathrm{Re}[v_\nu^{mn}v_\mu^{nm}](\delta(\omega-\Delta \omega_{nm})-\delta(\omega+\Delta\omega_{nm}))\right.\nonumber\\
&\left.-\mathrm{Im}[v_\nu^{mn}v_\mu^{nm}]\left(\mathcal{P}\frac{1}{\omega-\Delta\omega_{nm}}+\mathcal{P}\frac{1}{\omega+\Delta\omega_{nm}}\right)\right\},
\end{align}
where $\Delta\omega_{nm}=(\varepsilon_n-\varepsilon_m)/\hbar$ and $\Delta n_F^{mn}=n_F(\varepsilon_{m})-n_F(\varepsilon_{n})$ are the energy and distribution differences, respectively, between two states $|m\rangle$ and $|n\rangle$.
In threefold Hopf semimetals, the electronic energy $\varepsilon_m$ and the velocity component $v_\nu^{mn}$ are given in Eqs. (\ref{eq_energy}) and (\ref{eq_velocity}), respectively.
For the bulk material, the summation can be replaced by an integration in the wave number $\boldsymbol{q}$.
In the optical excitation process, the wave number is almost unchanged with a negligibly small momentum of the photon, and thus two electronic states $|m\rangle$ and $|n\rangle$ can be assumed to possess the same wave number.
Since the energy is isotropic and the velocity is scale free, the angle and amplitude integration can be performed separately,
\begin{align}
\sigma^{(1)}_{\nu\mu}(\omega)=&\frac{e^2}{\hbar}\sum_{\alpha<\alpha'}
\left\{\mathrm{Re}[A_{\nu\mu}^{\alpha\alpha'}]F^{\alpha\alpha'}_1(\omega)+\mathrm{Im}[A_{\nu\mu}^{\alpha\alpha'}]F^{\alpha\alpha'}_2(\omega)\right\},\label{eq_analytic_form}
\end{align}
where $\hat{\theta}$ represents the solid angle, with
\begin{align}
A^{\alpha\alpha'}_{\nu\mu}=\int d^2\hat{\theta}\ v_\nu^{\alpha\alpha'}v_\mu^{\alpha'\alpha},
\end{align}
and
\begin{align}
F^{\alpha\alpha'}_1(\omega)=&\frac{\pi}{\omega}\int^{q_c}_{0}\frac{dq}{(2\pi /a)^3}{q}^2\delta(\omega-\Delta\omega_{\alpha\alpha'}(q))\Delta n_F^{\alpha\alpha'}(q),\\
F^{\alpha\alpha'}_2(\omega)=&\frac{1}{\omega}\int^{q_c}_{0}\frac{dq}{(2\pi /a)^3}\mathcal{P}\frac{2\omega q^2\Delta n_F^{\alpha\alpha'}(q)}{(\Delta\omega_{\alpha\alpha'}(q))^2-\omega^2}.
\end{align}
Here, the cut-off wave number $q_c$ indicates the radius of the sphere in which the cone and flat band are preserved, and $a$ is introduced as a normalization factor for the integral.

The anisotropy of the optical conductivity is described by $A_{\nu\mu}^{\alpha\alpha'}$ which can be easily calculated by using Eq.  (\ref{eq_velocity}).
The non-zero components are given by
\begin{align}
A_{xx}^{+-}&=A_{yy}^{+-}=A_{zz}^{0\pm}=\frac{4\pi}{3}{v}^2,\\
A_{xx}^{0\pm}&=A_{yy}^{0\pm}={i\xi}A_{xy}^{0\pm}=\frac{2\pi}{3}{v}^2.
\end{align}
and the other components vanish due to the angle dependence.
Since the inter-cone component is absent for $A_{zz}^{\alpha\alpha'}$, the linearly polarized photon in the $z$ axis does not induce the electronic excitation between the two cone-shaped bands.
When the polarization axis is parallel to the $x$ or $y$ axis,  any electronic excitation is allowed among the three bands. 
The angle integrals also show the presence of optical Hall conductivity in the $xy$ plane, i.e., linearly polarized photons within the $xy$ plane induce an electric current perpendicular to the polarization axis.

At zero temperature, the analytic representation of the response function can be obtained.
In this condition, the Fermi distribution function can be replaced with a step function, $n_F(\varepsilon)\simeq\theta(\varepsilon_F-\varepsilon)$, in the calculation.
Thus, the optical response is non-zero if the photon energy is larger than the Fermi energy, $|\varepsilon_F|<\hbar\omega$, between the flat-band and a cone, and twice the Fermi energy,$|\varepsilon_F|<\hbar\omega$,  between two cones.
The analytic formula of optical conductivity is represented by
\begin{align}
\begin{aligned}
\sigma^{(1)}_{xx}(\omega)=&\frac{e^2}{h}\omega\frac{a^3}{6}\left(\frac{1}{2v}\theta(\hbar\omega-2\varepsilon_F)+\frac{1}{v}\theta(\hbar\omega-\varepsilon_F)\right),\\
\sigma^{(1)}_{zz}(\omega)=&\frac{e^2}{h}\omega\frac{a^3}{3v}\theta(\hbar\omega-\varepsilon_F),\\
\sigma^{(1)}_{xy}(\omega)
=&\xi\frac{e^2}{h}\frac{a^3}{3\pi{v}}\left\{(q_c-q_F)+\omega\log\left|\frac{1+\gamma_F}{1-\gamma_F}\right|\right\},
\end{aligned}\label{eq_analytic_formula}
\end{align}
with the Fermi wave number $q_F$ and $\gamma_F=\hbar\omega/vq_F$.
Here, the longitudinal conductivity is isotropic in the $xy$ plane, i.e., $\sigma^{(1)}_{yy}(\omega)=\sigma^{(1)}_{xx}(\omega)$.
In the third equation, it is assumed that the frequency is much smaller than the cut-off energy, i.e., $\hbar\omega/vq_c\ll1$.
This assumption works for investigating the characteristic property of the electronic structure around the point node.

The frequency dependence of the longitudinal optical conductivity shows the characteristic features of internal anisotropy in threefold Hopf semimetals. 
In the $z$ direction, the spectrum is similar to that of threefold point node semimetal\cite{Sanchez2019}, where the spectrum is zero for $\hbar\omega<\varepsilon_F$ and shows the linear dependence on the frequency for $\hbar\omega>\varepsilon_F$ with a step-like increment at $\hbar\omega=\varepsilon_F$.
In the case of a threefold point node semimetal, the optical conductivity shows the same spectrum in any direction but, in the threefold Hopf semimetal, the conductivity strongly depends on the polarization axis of the incident photon.
The optical conductivity within the $xy$-plane shows another characteristic feature attributed to the electronic state in the Hopf semimetal.
In the spectrum, there are two step-like increments found at $\hbar\omega=\varepsilon_F$ and $2\varepsilon_F$.
The first step and the second step correspond to the electronic excitation between the flat band and one cone,  and that between two cones, respectively.
The second one is a unique feature in a Hopf semimetal because the inter-cone excitation is prohibited in conventional threefold point node semimetals.

The optical Hall conductivity $\sigma_{\mu\nu}^{(1)}(\omega)$ can be observed in the $xy$-plane but it vanishes if the polarization axis or electric current is parallel to the $\boldsymbol{d}$-vector.
The analytic representation of $\sigma_{xy}^{(1)}(\omega)$ includes the frequency-independent term proportional to $q_c-q_F$. 
Although this term depends on the cut-off parameter $q_c$ explicitly, it does not mean that this term is an unrealistic one.
The frequency-independent term represents the conventional Hall current following the oscillating electric field.
The amplitude of this component decreases linearly with the Fermi energy and vanishes with it reaching $q_F=q_c$. 
The disappearance corresponds to that in association with full occupation of all electronic bands in a crystalline material.
In crystalline materials, the electronic states form bands bounded in the Brillouin zone and the summation of the Berry phase over all bands must be zero.

\subsection{Relation to symmetry}
In the previous section,  it was shown that the optical conductivity visualizes the internal anisotropy in the spectrum within the minimal model.
The anisotropic behavior in the spectrum is independent of the minimal model and directly attributed to the electronic property of the Hopf semimetal.
In this section, the number of step-like structure is discussed in terms of the three symmetries protecting the multifold point node.
The symmetries introduce a restriction on the electronic excitation responsible for the optical response represented by the general expression in Eq.\ (\ref{eq_response_function}).
Since there are three bands attaching at the point node, there are two possible excitation processes, the excitation from the lower cone to the upper cone, and that between the flat band and one cone. 
The former excitation process introduces a step-like structure at $\hbar\omega=2\epsilon_F$ and the latter produce another step-like structure at $\hbar\omega=\epsilon_F$ (see Figs.\ \ref{fig_delta_dependence} and \ \ref{fig_T_dependence}).
Theoretically, the absence of electronic excitation is described by the absence of off-diagonal velocity component $v_\mu^{mn}=\langle\psi_m|v_\mu|\psi_n\rangle$.
The off-diagonal component vanishes only for the $z$-direction, the parallel axis to the $d$-vector, between the upper and lower cones as shown in Eq.\ (\ref{eq_velocity}).

The prohibition of inter-cone excitation can be described by the symmetries protecting the multifold point node.
The chiral and $z$-axis mirror symmetries are relevant for the inter-cone excitation.
For the energy eigenstates, the chiral operator exchanges the electronic states, $|\psi_{\pm,\boldsymbol{q}}^\xi\rangle$, between the upper and lower cones,
\begin{align}
S|\psi_{\pm,\boldsymbol{q}}^\xi\rangle=c_S|\psi_{\mp,\boldsymbol{q}}^\xi\rangle,
\end{align}
with $c_S=1$ or $-1$ due to $S^2=I$.
On the other hand, the $z$-axis mirror operator $M_z$ causes electronic states to remain in the same band,
\begin{align}
M_z|\psi_{\pm,\boldsymbol{q}}^\xi\rangle=c_{M_z}^{\pm}|\psi_{\pm,\boldsymbol{q}}^\xi\rangle.
\end{align}
Here, the coefficient is the same, $c_{M_z}^+=c_{M_z}^-=1$ or $-1$, because the chiral and mirror operators are commutable $[S,M_z]$. 
Thus, in the presence of the two symmetries, the velocity operator $v_z$ never induces the inter-cone excitation because of
\begin{align}
\langle \psi^\xi_{+}|v_z|\psi^\xi_{-}\rangle&=c_{M_z}^2\langle \psi^\xi_{+}|(M_z^\dagger v_zM_z)|\psi^\xi_{-}\rangle\nonumber\\
&=\langle \psi^\xi_{+}|(-v_z)|\psi^\xi_{-}\rangle=0.
\end{align}
Therefore, the different number of steps in Eq. (\ref{eq_analytic_formula}) can be a signature of the internal anisotropy in threefold Hopf semimetals.

The absence of inter-cone excitation can also be observed in conventional threefold point node semimetals\cite{Sanchez2019,Habe2019} but the mechanism is different from that in threefold Hopf semimetals.
In the conventional threefold point node semimetals, electronic states can be described by a model of spin-1 particles and those in the upper and lower cones possess $\pm1$ spin angular momentum.
The difference in angular momentum leads to the prohibition of optical excitation  for any direction of the electric field.
In the threefold Hopf semimetals, on the other hand, the absence of inter-cone excitation is attributed to the symmetries protecting the point node.

\subsection{Numerical analysis}
In this section, the optical conductivity is numerically investigated including the effects of disorder and temperature.
The numerical analysis for the optical conductivity is performed by using Eq. (\ref{eq_response_function}).
The non-zero temperature changes the distribution function $n_F(\varepsilon)=(1+\exp[(\varepsilon-\varepsilon_F)/(k_BT)])^{-1}$ and the disorder is considered a reduction of the electronic relaxation time in the calculation where the relaxation time $\tau$ is introduced by replacing $\delta$ with $\hbar/(2\tau)$ in Eq. (\ref{eq_response_function}).
The summation is performed over the band index and the wave number within the sphere possessing a diameter $q_c$.
Since the electronic states are scale free in the wave number space, the Fermi energy $\varepsilon_F$ is set to unity and the other parameters, $k_BT$, $\hbar vq_c$, and $\delta=\hbar/2\tau$, are normalized by $\varepsilon_F$. 
The cut-off energy is ten times the Fermi energy, $\hbar vq_c=10\varepsilon_F$.
In the numerical results, some constants, $k_B$, $\hbar$, and $v$, are set to unity for simplicity.
The Drude peak, the electronic conductivity without inter-band excitation, is omitted in the calculation.

\begin{figure}[htbp]
\begin{center}
 \includegraphics[width=70mm]{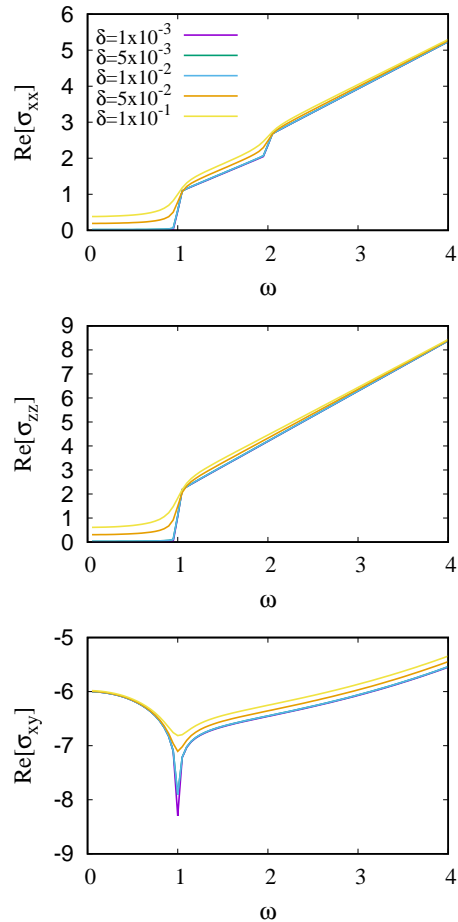}
\caption{The frequency dependence of the optical conductivity of a Hopf semimetal with changing electronic relaxation time $\tau=\hbar/2\delta$. The other parameter is fixed as $T=1\times10^{-3}$.
 }\label{fig_delta_dependence}
\end{center}
\end{figure}
The effect of disorder is analyzed numerically as shown in Fig.\ \ref{fig_delta_dependence}.
The longitudinal optical conductivity shows a linear dependence on the frequency larger than the Fermi energy in  both the $x$ and $z$ directions, where the longitudinal conductivity in the $y$ direction is equivalent to that in the $x$ direction.
In the $x$ and $y$ directions, another step can be observed in the spectrum at twice the Fermi energy and it does not appear in that of the $z$ direction. 
The anisotropy is consistent with the analytic results in the clean limit in Eq. (\ref{eq_analytic_formula}).
These steps are broadened with an increase in $\delta$, the decrease in relaxation time,but they can be observed for $\delta=1\times10^{-1}$, which is equal to 10meV for $\varepsilon_F=100$meV.
The optical Hall conductivity $\sigma_{xy}^{(1)}$ shows a negative peak structure around $\omega=\varepsilon_F$. 
It is represented by the second term of $\sigma_{xy}^{(1)}$ in Eq. (\ref{eq_analytic_formula}).
The amplitude decreases with the frequency and it is also consistent with the analytic result.
\begin{figure}[htbp]
\begin{center}
 \includegraphics[width=70mm]{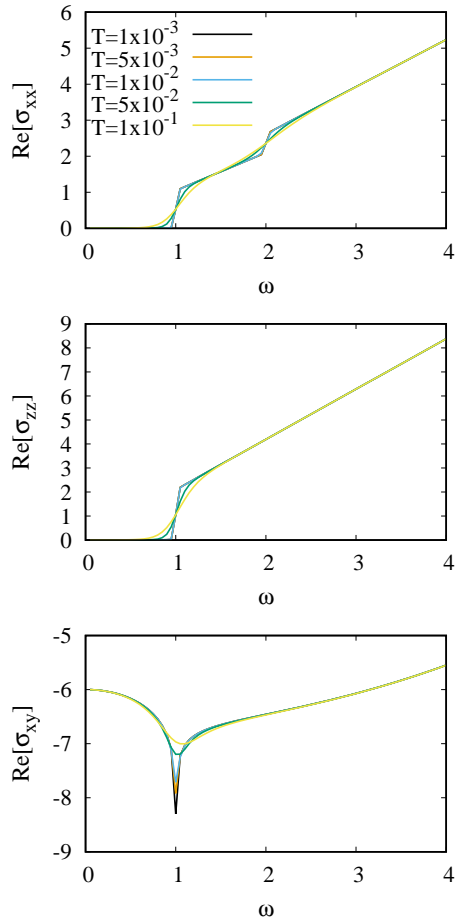}
\caption{The frequency dependence of the optical conductivity of a Hopf semimetal with changing temperature $T$. The other parameter is fixed as $\delta=1\times10^{-3}$.
 }\label{fig_T_dependence}
\end{center}
\end{figure}
The temperature dependence is presented in Fig.\ \ref{fig_T_dependence}.
The characteristic structures in the spectrum change to be smooth but they can be observed even at non-zero temperature.

These results by using the minimal model show the ideal spectra of optical conductivity.
Since the spectra are modified in a realistic material due to the more complex electronic structure, possible corrections are discussed in what follows.
Firstly, the linear and flat bands are not preserved in the whole Brillouin zone.
The characteristic spectrum can be observed for the frequency within the excitation energy among the bands which can be described by Eq.\ (\ref{eq_energy}).
Secondly, there can be other Fermi pockets where electronic states cannot be described as multifold fermions.
However, in the case of  the optical response, such Fermi pockets do not contribute to the low-frequency spectrum unless low-energy excited states accompany them.
In practice, the optical conductivity of conventional multifold fermion is reproduced in realistic materials of a multifold point node semimetal with such additional Fermi pockets\cite{Sanchez2019, Habe2019}.

\section{conclusion}\label{sec_conclusion}
In this paper, a theoretical analysis of the optical conductivity of threefold Hopf semimetals is presented by using a simplified theoretical model possessing an isotropic nodal band structure and anisotropy represented by the $d$-vector in the electronic states.
In the clean and zero-temperature limit, the analytic representations of the optical conductivity are obtained in Eq. (\ref{eq_analytic_formula}).
The theoretical calculations reveal that the effect of internal anisotropy can be observed in the spectrum of the optical conductivity.
The spectrum of the longitudinal optical conductivity shows a linear dependence on the frequency and step-like behaviors at some frequencies.
For an optical polarization axis perpendicular to the $d$-vector, two steps can be found at the frequency corresponding to the Fermi energy and twice the Fermi energy.
On the other hand, for the polarization axis parallel to the $d$-vector, the step at twice the Fermi energy vanishes and a single step is found in the spectrum.
The qualitative anisotropy is attributed to the prohibition of inter-cone electronic excitation due to chiral and mirror symmetries protecting the point node.
Moreover, the optical Hall conductivity can be observed only if the current and electric field are perpendicular to the $d$-vector. 
The numerical calculations show that the anisotropic behaviors are stable even with the non-zero temperature and electronic relaxation time.

This work was supported by JSPS KAKENHI Grant No. JP20K05274.

\bibliography{Topological_SM}

\end{document}